\documentclass[copyright,creativecommons]{eptcs}
\providecommand{\event}{TARK 2021} 

\usepackage[utf8]{inputenc}
\usepackage{amsmath}
\usepackage{amssymb}
\usepackage{amsthm}
\usepackage{stmaryrd}
\usepackage{tikz}
\usepackage{enumerate}
\usepackage{soul}
\usepackage{centernot}
\usepackage{enumitem}
\usepackage{xcolor}
\usepackage{multirow}

\usepackage{amsfonts}
\DeclareFontFamily{U}{mathb}{\hyphenchar\font45}
\DeclareFontShape{U}{mathb}{m}{n}{
      <5> <6> <7> <8> <9> <10> gen * mathb
      <10.95> mathb10 <12> <14.4> <17.28> <20.74> <24.88> mathb12
}{}
\DeclareSymbolFont{mathb}{U}{mathb}{m}{n}
\DeclareMathSymbol{\llcurly}{3}{mathb}{"CE}
\DeclareMathSymbol{\ggcurly}{3}{mathb}{"CF}

\usepackage{graphicx}
\graphicspath{ {./} }
\usepackage{subcaption}

\begin{document}

\title{Knowledge from Probability}
\author{Jeremy Goodman
\institute{School of Philosophy\\University of Southern California, USA}
\email{jeremy.goodman@usc.edu}
\and
Bernhard Salow
\institute{Faculty of Philosophy\\University of Oxford, UK}
\email{bernhard.salow@philosophy.ox.ac.uk}
}
\def\titlerunning{Knowledge from Probability}
\def\authorrunning{J. Goodman \& B. Salow}

\maketitle

\begin{abstract}
    We give a probabilistic analysis of inductive knowledge and belief and explore its predictions concerning knowledge about the future, about laws of nature, and about the values of inexactly measured quantities. The analysis combines a theory of knowledge and belief formulated in terms of relations of comparative normality with a probabilistic reduction of those relations. It predicts that only highly probable propositions are believed, and that many widely held principles of belief-revision fail.
\end{abstract}

How can we have knowledge that goes beyond what we have observed -- knowledge about the future, or about lawful regularities, or about the distal causes of the readings of our scientific instruments? Many philosophers think we can't. 
Nelson Goodman, for example, disparagingly writes that ``obviously the genuine problem [of induction] cannot be one of attaining unattainable knowledge or of accounting for knowledge that we do not in fact have'' \cite[p. 62]{Good55}. Such philosophers typically hold that the best we can do when it comes to inductive hypotheses is to assign them high probabilities. Here we argue that such pessimism is misplaced. We give a purely probabilistic analysis of inductive knowledge and (rational) belief, and motivate it by drawing out its attractive predictions about a range of cases.

Our analysis builds on two recent strands of research. The first strand is the idea that knowledge and (rational) belief can be analyzed in terms of a notion of \emph{normality}: among the possibilities that are compatible with an agent's evidence, their knowledge rules out those that are sufficiently less normal than their actual circumstances, and their beliefs rule out those that are sufficiently less normal than some other evidential possibilities.\footnote{\label{fn:normos} See \cite{Stal06, Stal15,Stal19, Good13, Grec14, Duta16, GoodSalo18, GoodSaloEpis, Cart19, LittDuta20, BeddPaveFC, CartGoldms, LoetMS, GoldHawtContam} for related ideas about knowledge, \cite{Smit10, Smit16, Smit17, Smit18, Smit18logic} for related ideas about justified belief, and \cite{KrauLehmMagi90, Maki93} and references therein for related ideas about non-monotonic reasoning.}
The second strand is that what a person knows or believes is always relative to a contextually supplied question.\footnote{See \cite{SchaSzab14, Yabl14, HolgMSconstraint} for precedents in the case of knowledge, and \cite{Leit14Stability, Yalc18, HolgMS, BlumLedeFC} for precedents in the case of belief.} Our guiding observation is this: \emph{there is a natural way to define the comparative normality of evidential possibilities in terms of the probabilities of the answers to a  question}. This fact allows us to give an analysis of knowledge and belief in terms of probability and evidence, two notions that even skeptics about inductive knowledge typically accept. 

Here is our plan. We begin by presenting a version of the theory of knowledge and belief in terms of comparative normality that we have defended elsewhere 
\cite{GoodSalo18, GoodSaloEpis}. We next explain how comparative normality can be reduced to evidential probability in a question-relative way. We then use this framework to model knowledge and belief about ongoing chancy processes (section~\ref{sec:flipping}), lawful regularities (section~\ref{sec:laws}), multiple independent subject matters (section~\ref{sec:racing}), and the values of quantities measured using instruments that are subject to random noise (section~\ref{sec:clock}).\footnote{See \cite{DGH14, Baco14, BacoFC, Smit17, Smit18, GoodSalo18, GoodSaloEpis} for recent discussion of the kind of cases in $\S\S$3-5, and \cite{Will11, Will13, Will14,Will20, CoheCome13, Good13, Weat13, Stal15, Cart19, DutaRoseFC, CartGoldms, GoodSaloEpis} for recent discussion of the kind of cases in $\S6$.} 
As these case studies will illustrate, the framework is conservative in its synchronic predictions but revisionist in its diachonronic ones: what an agent knows and believes is closed under entailment and always has a high probability of being true, but getting new evidence can lead to changes in what one believes that violate widely endorsed principles about belief-revision. Three appendices explain ways in which the framework can be extended in order to model a wider range of cases of inductive knowledge.

\section{The Normality Framework}\label{sec:framework}

Our models of knowledge and belief will be a version of Hintikka semantics for a single agent.\footnote{We can model the knowledge/beliefs of $n$ agents by generalizing clause 2 of the definition of a normality structure so $\mathcal{E}\subseteq\mathcal{P}(S)^n$ (giving the possible patterns of bodies of evidence among the agents) and modify the remaining definitions accordingly.} Both knowledge and belief are given by accessibility relations, in the sense that an agent knows/believes $p$ in a world $w$ iff $p$ is true in all worlds epistemically/doxastically accessible from $w$. What is distinctive of the framework is how these accessibility relations are defined in terms of other relations between worlds, encoding the agent's evidence and worlds' comparative normality. 

Let a \emph{normality structure} be a tuple $\langle S, \mathcal{E},W,\succcurlyeq,\ggcurly\rangle$ such that:
\begin{enumerate}
    \item $S$ is a non-empty set (of \emph{states}),
    \item $\mathcal{E}\subseteq\mathcal{P}(S)\backslash \{\emptyset\}$ (the \emph{possible bodies of evidence})
    \item $W= \{\langle s,E\rangle: s\in E\in \mathcal{E}\}$ (the set of \emph{(centered) worlds}),
    \item $\succcurlyeq$ is a preorder on $W$ (read `$w\succcurlyeq v$' as `$w$ is \emph{at least as normal as} $v$'),
    \item $\ggcurly$ is a well-founded relation on $W$ (read `$w\ggcurly v$' as `$w$ is \emph{sufficiently more normal than} $v$'), such that, for any worlds $w_1,w_2,w_3,w_4$:\begin{enumerate}
        \item If $w_1\ggcurly w_2$, then $w_1\succcurlyeq w_2$;
        \item If $w_1\succcurlyeq w_2 \ggcurly w_3 \succcurlyeq w_4$, then  $w_1 \ggcurly w_4$.
        \end{enumerate}
\end{enumerate}
The intuitive idea behind modeling worlds as state/set-of-state pairs is that we are only modelling the agent's knowledge and beliefs about a certain subject matter -- the state of the world -- and for this purpose we may idealize and treat worlds as individuated by the state of the world together with the agent's evidence about the state of the world, modelled as the set of states compatible with their evidence. As we will understand it, a person's evidence is a subset of their knowledge, and hence is true; this is why in any world the actual state is a member of the set of states compatible with the agent's evidence. We use $R_e$ to denote the function mapping each world to the set of worlds that are \emph{evidentially accessible} from it, in the sense of being compatible with the agent's evidence:
\begin{itemize}
    \item[] $R_e(\langle s,E\rangle):=\{\langle s',E\rangle: s'\in E\}$
\end{itemize}

Next, we define a function $R_b$ for \emph{doxastic accessibility}, characterizing the set of worlds compatible with what the agent believes in any given world. The idea is that what the agent believes goes beyond what is entailed by their evidence: the doxastic possibilities are those evidential possibilities that are not sufficiently less normal than any other. Formally,
\begin{itemize}
    \item[] $R_b(w):=\{v\in R_e(w):\neg(\exists u\in R_e(w):u\ggcurly v)\}$
\end{itemize}
Finally, we define a function $R_k$ for \emph{epistemic accessibility}, characterizing the set of worlds compatible with what the agent knows in any given world. There are two natural definitions here: one for those who follow Stalnaker \cite{Stal06,Stal15,Stal19} 
in thinking that epistemic accessibility is a transitive relation (in which case knowing $p$ entails knowing that you know $p$), and another for those who follow Williamson \cite{Will00,Will11,Will13b,Will14} 
in thinking that epistemic accessibility cannot be transitive because knowledge requires a margin for error. The difference concerns whether worlds that are doxastically inaccessible should be epistemically accessible when they are less normal but not sufficiently less normal than actuality: the Stalnakerian answers ``no'' (the agent knows those worlds don't obtain), while the Williamsonian answers ``yes'' (for all the agent knows, those worlds obtain). Formally, these answers correspond to the following respective definitions:
\begin{itemize}
    \item[]$R_{k_{\text{}}}(w):=R_b(w)\cup\{v\in R_e(w): v\succcurlyeq w\}$ \hspace{6cm}(Stalnakerian)
    \item[] \vspace{-.5cm}$R_k(w):=R_b(w)\cup\{v\in R_e(w):v\succcurlyeq w\}$ $\cup$ $\overbrace{\{v\in R_e(w):w\succcurlyeq v\wedge\neg(w\ggcurly v)\}}^{\text{the margin for error}}$ \hspace{.5cm}(Williamsonian)
 \end{itemize}
The results we will be exploring in this paper don't depend on which definition of knowledge we adopt.\footnote{Note that the Williamsonian definition is equivalent to the much simpler definition $R_k(w):=\{v\in R_e(w):\neg(w\ggcurly v)\}$ in normality structures where $\succcurlyeq$ is a \emph{total} preorder on $E$ for all $E\in\mathcal{E}$, which includes all normality structures generated from probability structures in the way described below; see \cite{GoodSaloEpis} (which considers a slightly more complicated Williamsonian definition to ensure that the set of epistemically accessible evidential possibilities is closed under $\succcurlyeq$).} 
Although we will remain neutral on which definition is preferable by focusing mainly on belief, we believe that the normality framework is recommended in large part by its ability to integrate an anti-skeptical theory of knowledge with a non-trivial theory of inductive belief; we make this case at greater length in \cite{GoodSaloEpis}.

The framework also allows us to model the dynamics of knowledge and belief \emph{about the state of the world} in response to new evidence about the state of the world. To make this idea precise, we first introduce the projection functions $\pi_i$ such that $\pi_i(\langle x_1,\dots,x_n\rangle)=x_i$. We then define which \textit{states} are accessible from a world $w$ as $\mathcal{R}_*(w)=\{\pi_1(v):v\in R_*(w)\}$ where $*\in\{e,b,k\}$. (So, e.g., $\mathcal{R}_e(w)=\pi_2(w)$.) For any set of states $p$ and pair of worlds $w$ and $v$, we say that $v$ \emph{is the result of discovering $p$ in} $w$ iff $\pi_1(v)=\pi_1(w)$ and $\pi_2(v)=p\cap\pi_2(w)$. Although $R_b(w)\cap R_b(v)=\emptyset$ whenever $v$ is the result of (non-trivially) discovering $p$ in $w$ (non-trivially in the sense that $w\neq v$), the dynamics relating $\mathcal{R}_b(w)$ and $\mathcal{R}_b(v)$ (and $\mathcal{R}_k(w)$ and $\mathcal{R}_k(v)$) are more interesting, as we will explore below. Note that, unlike standard models of belief-revision, there may be no $v$ that is the result of discovering $p$ in $w$ -- for example, if $p$ is incompatible with the state of the world in $w$. And since only truths can be discovered, normality structures allow us to easily model iterated discoveries (in contrast to formally similar models of theories of belief-revision like AGM \cite{AGM85}, first developed in \cite{Grov88}, which do not handle iterated belief-revision).\footnote{Normality structures are also related to models in dynamic epistemic logic, since instead of modeling accessibility as a relation between worlds, we could equivalently treat it as family of relations between states indexed by bodies of evidence, much like how accessibility is relativized to propositions in dynamic epistemic logic. Another formal precedent is \cite{DabrMossPari96}, in which formulas are evaluated relative to a pair of a world and a set of worlds containing it. (Thanks to Ayb\"{u}ke \"{O}zg\"{u}n for drawing our attention to this work.) Note that what ``world'' the agent is in changes as they get new evidence; those who prefer to reserve the word ``world'' for something unchanging can substitute ``situation'', ``case'', or ``centered world''.}$^,$\footnote{The framework presented here 
sides with \cite{GoodSaloEpis} over \cite{GoodSalo18} by treating normality relations as holding between worlds rather than states. But it sides with \cite{GoodSalo18} by presupposing that worlds can be factored into state/evidence pairs, which in turn implies that evidential accessibility is an equivalence relation. Appendix \ref{app:Generalizations} explains how to modify the definitions below in order to model scenarios in which evidential accessibility is not an equivalence relation.}

\section{Reducing normality to probability}\label{sec:reduction}

Appealing to notions of comparative normality (or comparative plausibility) is, by now, a familiar idea in theorizing about knowledge and belief. The main advance of this paper is to explore the consequences of an analysis of these notions in terms of the result of conditioning a prior probability distribution on the agent's evidence. 

Let us begin with the at-least-as-normal relation $\succcurlyeq$. An initially attractive idea is that $w \succcurlyeq v$ iff $w$ is at least as probable as $v$. Unfortunately, this simple proposal faces a number of problems. For example, the probability of a world depends on how finely we individuate worlds in our model in ways that intuitively shouldn't make a difference to what an agent knows or believes; also, natural ways of individuating worlds often make them all have the same probability, thereby trivializing inductive knowledge and belief. 
For these reasons, as well as others explained in section~\ref{sec:racing}, we will model knowledge and belief as relative to a contextually supplied question about the state of the world. Doing so allows for a more robust characterization of normality in terms of probability, as we will now explain. 

\begin{samepage} Let a \emph{probability structure} be a tuple $\langle S,\mathcal{E},W,Q,P,t\rangle$ such that:
\begin{enumerate}
    \item $S,\mathcal{E},W$ satisfy clauses 1-3 of the definition of normality structures,
    \item $Q$ (the \emph{question}) is a partition of $S$,
    \item $P$ (the \emph{prior}) is a probability distribution over 
    $S$ such that $P(q|E)$ is defined for all $q\in Q$ and $E\in \mathcal{E}$,
    \item $t\in [0,1]$ (the \emph{threshold}).
\end{enumerate} 
\end{samepage}
We will now explain how to generate a normality structure from a probability structure. We identify the normality of a world with the evidential probability at that world of the true answer to $Q$ at that world. Formally, letting $[s]_Q$ (the \emph{answer to} $Q$ in $s$) be the cell of $Q$ containing $s$, $P_w$ (the \emph{evidential probability} at $w$) be $P(\cdot|\pi_2(w))$, and $\lambda(w)$ (the \emph{likeliness} of $w$) be $P_w([\pi_1(w)]_Q)$, we adopt the following definition of one world being at least as normal as another:

\begin{itemize}
    \item[] \textsc{normality as likeliness}: $w\succcurlyeq v:=\lambda(w)\geq \lambda(v)$ and $v\in R_e(w)$
\end{itemize}

Next, let the \emph{typicality} of a world be the evidential probability, at that world, that things are no more normal than they are at that world: formally, $\tau(w)=P_w(\{\pi_1(v): w\succcurlyeq v$ and $v\in R_e(w)\})$. We will adopt the following definition of one world being sufficiently more normal than another:
\begin{itemize}
    \item[] \textsc{sufficiency}: $w\ggcurly v:=1-\frac{\tau(v)}{\tau(w)} \geq t$ and $v\in R_e(w)$

\end{itemize}
It is easy to verify that, so defined, $\langle S,\mathcal{E},W,\succcurlyeq,\ggcurly\rangle$ is a normality structure.\footnote{
The requirement that $\succcurlyeq$ and $\ggcurly$ only relate evidentially accessible worlds is needed to validate conditions 5a and 5b of the definition of a normality structure, since worlds with the same likeliness but different evidence can have different typicality. For example, let $S=\{1,\dots,7\}, \mathcal{E}=\{\{1,2,3\},\{4,5,6,7\}\},Q=\{\{s\}:s\in S\},P(1)=P(2)=P(4)=.2;P(3)=P(5)=P(6)=P(7)=.1; t=.5$: Let $w=\langle 3,\{1,2,3\}\rangle$ and $v=\langle 5,\{4,5,6,7\}\rangle$. $\lambda(w)=\lambda(v)=.2$, but $\tau(w)=.2\neq\tau(v)=.6$.
} 
In this normality structure, what the agent believes about the state of the world is the strongest disjunction of answers to $Q$ that (i) includes the most probable answers, (ii) includes all answers at least as probable as any it includes, and (iii) has total probability at least $t$.\footnote{Compare the theory of belief in \cite{HolgMS}, which implies (i) and a slight weakening of (ii) (with ``more probable'' in place of ``at least as probable''), but not (iii). Rather than seeing this view as a competitor to ours, we prefer to see it as concerned with a weaker notion of belief that is less closely tied to knowledge. \cite{HongDiss} has independently developed a theory of knowledge and belief very close to the one presented here, which he applies to the preface and St. Petersburg paradoxes.} (This will also be what the agent knows about the state of the world if one of the most probable answers to $Q$ is true.) \textsc{normality as likeliness} ensures (i) and (ii), while \textsc{sufficiency} ensures (iii), which may be more precisely stated as follows:\footnote{\textsc{sufficiency} does make some somewhat surprising predictions. Suppose Alice holds 999 tickets in a fair lottery and each of the million other entrants holds 1000 tickets. Assume $P$ gives each ticket an equal probability of being chosen given the setup, that the question $Q$ is \emph{who will win}, and the threshold $t= .99$, and consider worlds in which the agent's evidence is that this is the setup. If Alice will lose, can the agent know this? \textsc{sufficiency} predicts they can, which seems odd. (This might not seem so odd at first, since we don't deny the agent can know this relative to a different question $Q'$ = \emph{how many tickets does the winner have}. To bring out the oddity, we can modify the example by adding another loser Bob who holds 1001 tickets; it seems odd that the agent would know that Alice will lose but not that Bob will lose.)  To avoid this prediction, we might add a requirement that a world is sufficiently more normal than another only if it is also sufficiently more \emph{likely}. (Compare the discussion of having no ``\emph{appreciably} stronger reason'' (emphasis ours) to believe one entrant to a lottery will win compared to any other in \cite[p.16]{Hawt04}.) We might implement this idea by adopting the stronger principle:\begin{itemize}\item[] \textsc{sufficiency+}: $w\ggcurly v:=1-\frac{\tau(v)}{\tau(w)} \geq t$ and $v\in R_e(w)$ and $1-\frac{\lambda(v)}{\lambda(w)}\geq t$\end{itemize} This definition also determines a normality structure; while we actually find it more attractive than \textsc{sufficiency}, we will ignore it in the main text to simplify our presentation.
} 
\begin{itemize}
    \item[]\textsc{threshold}:  $P_w(\mathcal{R}_b(w))\geq t$ for all $w\in W$.
\end{itemize}

\section{Inductive knowledge about the future}\label{sec:flipping}

In this section and the next we'll review two cases of inductive knowledge involving coin flips from Dorr, Goodman and Hawthorne \cite{DGH14} to help illustrate the framework. The first case illustrates the possibility of inductive knowledge about the future: 
\begin{quote} \textbf{Flipping for Heads}: A coin flipper will flip a fair coin until it lands heads. Then he will flip no more.
\end{quote}
Assume that the agent's evidence entails that this is the setup, and that they are watching the experiment unfold. 
We can model the case using the probability structure in which $S=\{1,2,\dots\}$, $\mathcal{E}=\{\{n,n+1,n+2,\dots\}: n\in S\}$, $P$ is the probability function such that $P(\{n\})=2^{-n}$ for all $n$, $Q=\{\{s\}:s\in S\}$, and $t=.99$. Intuitively, $n$ is the state in which the coin lands heads on the $n$th flip, the possible bodies of evidence are those compatible with having watched some initial sequence of zero or more flips that all landed tails, and $Q$ is the question \emph{on what flip does the coin land heads}. In the normality structure generated from this probability structure, $\langle n,E\rangle\succcurlyeq \langle m,E\rangle$ iff $n\geq m$, and $\langle n,E\rangle\ggcurly \langle m,E\rangle$ iff $n\geq m+7$.\footnote{$\tau(\langle n,E\rangle)=2^{1-n}$, and $1-2^{-6}<.99<1-2^{-7}$, so  $1-\frac{\tau(\langle m,E\rangle)}{\tau(\langle n,E\rangle)}\geq .99$ iff $n\geq m+7$.} So if the agent has seen the coin land tails $x$ times, what they believe is that it will land heads within the next 7 trials -- i.e., on trials $x+1$ to $x+7$. 
The predictions about knowledge match those of \cite{DGH14}, on the Williamsonian definition of epistemic accessibility, and those of \cite{GoodSalo18}, on the Stalnakerian definition. 

Notice that this model involves a kind of non-monotonic belief revision (which is prohibited by AGM). Let $w=\langle 2,\{1,2,\dots\}\rangle$, $v=\langle 2,\{2,3,\dots\}\rangle$, and $p=\{2,3,\dots\}$. So $v$ is the result of discovering $p$ in $w$. At the start of the experiment, the agent is in $w$; after the first trial, the coin lands tails and the agent is in $v$ (and, unbeknownst to them, the coin is about to land heads). $\mathcal{R}_b(w)=\{1,\dots,7\}$, which is compatible with $p$, yet $\mathcal{R}_b(w)\cap p\neq\mathcal{R}_b(v)=\{2,\dots,8\}$. Discovering something compatible with their prior beliefs leads the agent to give up some those beliefs. We think this is exactly the right prediction.\footnote{Even more surprising behavior is possible if we modify the case by allowing the agent to get partial information about the result of the experiment after it is over, so that $\{1,\dots,7\}\in\mathcal{E}$. Consider $w=\langle 1,\{1,2,\dots\}\rangle$, $v=\langle 1,\{1,\dots,7\}\rangle$, and $p=\{1,\dots,7\}$. $\mathcal{R}_k(w)=\mathcal{R}_b(w)=p$ and $v$ is the result of discovering $p$ in $w$, yet $\mathcal{R}_k(v)=\mathcal{R}_b(v)=\{1,\dots,6\}$: you can gain new beliefs and knowledge by discovering (i.e., gaining \emph{evidential} knowledge of) something you already (inductively) knew.}

\section{Inductive knowledge of laws}\label{sec:laws}

The second case from \cite{DGH14} provides a simple model of inductive knowledge of lawful regularities:
\begin{quote} \textbf{Heading for Heads}: You know a bag contains two coins: one fair, one double-headed. Without looking, you reach in and select a coin. You decide to flip it 100 times and observe how it lands. 
\end{quote}
If inductive knowledge of lawful regularities is ever possible, it should be possible here: that is, you can learn that the coin is double headed by seeing it land heads 100 times in a row. Our framework predicts this. Consider the probability structure with $2^{100}+1$ states, encoding the pattern of heads and tails and whether the coin is fair or double-headed. Let $c$ be the state where the coin is fair but lands heads every time by coincidence, and $d$ be the state where it is double-headed. For every state there are two worlds, corresponding to your evidence before and after flipping, so $\mathcal{E}=\{S\}\cup\{c,d\}\cup\{\{s\}:s\in S\backslash\{c,d\}\}$. $P(\{d\})=.5$ and  $P(\{s\})=.5^{101}$ for $s\in S\backslash\{d\}$. $Q=\{\{s\}:s\in S\}$ and $t=.9999999$. As desired, this structure predicts that, after seeing the coin land heads 100 times, you can know it is double headed.\footnote{The theory similarly predicts that one can come to know how many sides of a die are painted red by observing the outcomes of a sufficient number of rolls of the die. It thus explains not only knowledge of law-like generalizations, but also knowledge of the objective chances associated with different physical processes.}

Notice that, although $\langle c,\{c,d\}\rangle\not\in R_k(\langle d,\{c,d\}\rangle)$, nevertheless $\langle c,S\rangle\in R_k(\langle d,S\rangle)$.\footnote{$1-\frac{\tau(\langle c,\{c,d\}\rangle)}{\tau(\langle d,\{c,d\}\rangle)}=\frac{2^{100}}{2^{100}+1}\geq.9999999$; but $1-\frac{\tau(\langle c,S\rangle)}{\tau(\langle d,S\rangle)}=.5<.9999999$.} In other words: although you know the coin is double-headed after seeing it land heads every time, nevertheless, before flipping it, for all you knew the coin was fair and about to land heads every time by coincidence. This fact highlights a notable feature of the present framework that departs from our models in \cite{GoodSalo18}: the comparative normality of two worlds is not a function only of those worlds' underlying states, but also depends on the agent's evidence. (This is perhaps more intuitive if we gloss $\succcurlyeq$ and $\ggcurly$ as relations of comparative \emph{plausibility}: what hypotheses about the state of the world are more or less plausible depends on what your evidence about the state of the world is.) The case also illustrates some more extreme departures from AGM-style dynamics of knowledge and belief, since discovering $p$ (that the coin landed heads every time) can allow one to know $q$ (that it isn't fair) even if, prior to the discovery, $p\wedge\neg q$ was an epistemic possibility. While such behavior is unfamiliar (and claimed in \cite{DGH14} to be impossible), we again submit that in this case it is a plausible prediction.\footnote{Note that the fact that, for all you know at the outset, the coin is fair and will land heads every time by coincidence, depends on the choice of question. Consider instead $Q'=$ \textit{is the coin fair and how many times will it land heads}. Relative to this question, you do know at the outset that, if the coin is fair, it won't land heads every time. This is because the change in question from $Q$ to $Q'$ changes the typicality of $\langle c,S\rangle$ from $.5$ to $.5^{-100}$ -- this is because fair-and-all-heads is less normal than all other states relative to $Q'$ except for fair-and-all-tails (which is equally normal).}

\section{Inductive knowledge about multiple subject matters}\label{sec:racing}

Theories of inductive knowledge that accept \textsc{threshold} face a well-known challenge. Assuming inductive knowledge is possible at all, we should be able to know propositions whose evidential probability is less than 1. Now consider many independent subject matters about which we have such inductive knowledge, and the conjunction of everything we know concerning any one of these subject matters. If knowledge is closed under conjunction, as Hintikka semantics predicts, it follows that we know this conjunction. But if we are pooling knowledge across enough independent subject matters, then this conjunction is liable to have low evidential probability, even if the evidential probaility of each of its conjuncts is high, thereby violating \textsc{threshold}.

Our proposal responds to this challenge by maintaining that ``know'' (and ``believe'') are context-sensitive, with the question $Q$ being the relevant parameter of context-sensitivity. For any given question, knowledge relative to that question is closed under conjunction. But there need be no single question relative to which an agent knows every proposition that they know relative to some question or other, and hence no question relative to which they know the conjunction of all such propositions.
We will illustrate this aspect of the proposal using the following case we discuss in \cite{GoodSalo18}:
\begin{quote} \textbf{Racing for Heads}: Each of $n$ coin flippers has a fair coin. Each will flip their coin until it lands heads.
\end{quote}
If what the agent knows/believe about each coin flipper is like what they know/believe about the single coin flipper in \textbf{Flipping for Heads}, then as $n$ grows the totality of what they know/believe can have arbitrarily low probability, violating \textsc{threshold}.\footnote{\cite{Will09} defends the existence of such \textsc{threshold} violations; \cite{GoodSalo18} shows how they can be modeled using normality structures.} We will now explain our preferred treatment of the case, in terms of probability structures. We will show how its predictions about the agent's knowledge and beliefs depend on the contextually supplied question. 

For concreteness, we will investigate the version of the case with 10 coin flippers. The probability structure is the one defined in the obvious way like in \textbf{Flipping for Heads}, with states modeled as sequences of 10 positive integers, indicating how many trials it will take for each of the 10 coins to land heads, and 
$P$ corresponding to the chances of various outcomes prior to the experiment. 

What about $Q$? In reflecting on \textbf{Racing for Heads} a number of natural questions suggest themselves. Some of these are the $10$ different questions of the form \emph{how many times will this particular coin land heads}. Relative to any such question, your knowledge is exactly like that in \textbf{Flipping for Heads} with respect to this particular coin, and trivial concerning every other coin. But other natural questions include (i) \textit{what will the exact outcome of the whole experiment be}; (ii) \textit{what will the shape of the outcome be} -- that is, how many coins will take how long to land (the exact outcome up to isomorphism); (iii)  \textit{how many total tails will there be in the experiment as a whole};  (iv) \textit{how long will it be before all the coins have landed heads}; and (v) \textit{how many of the coins will ever land heads at the same time}. For each of these question, we can ask what you know and believe about a number of issues, such as how many total tails there will be, how long the experiment as a whole will last, and whether all the experiments will end on the same flip (a claim labelled `same end' below). The table below records what the agent believes at the start of the experiment for different choices of $Q$, for thresholds $t=.75$ and $t=.95$. This will also be what the agent knows in the most normal worlds. 

\vspace{10pt}\hspace{1.2cm}
\noindent \begin{tabular}{|p{2.2cm}||p{2.75cm}|p{.35cm}|p{.5cm}p{.5cm}p{.8cm}p{.8cm}p{1cm}|}
\hline
\hspace{.9cm}$Q$&which worlds are most normal&$t$&min tails&max tails&min trials&max trials&same end?\\\hline\hline
\multirow{ 2}{2.4cm}{(i) exact outcome}& \multirow{ 2}{2.5cm}{all coins land heads first time}&.75&0&13&1&14&maybe\\
&&.95&0&18&1&19&maybe\\\hline
\multirow{ 2}{2.4cm}{(ii) outcome shape}& \multirow{ 2}{2.5cm}{$6 \times 1$ flip, $3 \times 2$ flips, $1\times 3$ flips}&.75&1&15&2&8&no\\
&&.95&0&22&1&12&maybe\\\hline
\multirow{ 2}{2.4cm}{(iii) how many total tails}& \multirow{ 2}{2.5cm}{8 or 9 total tails [tied]}&.75&5&14&1&15&maybe\\
&&.95&2&18&1&19&maybe\\\hline
\multirow{ 2}{2.4cm}{(iv) how long until over}& \multirow{ 2}{2.5cm}{ends on 4$^{\text{th}}$ trial}&.75&2&50&3&6&maybe\\
&&.95&1&70&2&8&maybe\\\hline
\multirow{ 2}{2.4cm}{(v) how many end together}& \multirow{ 2}{2.5cm}{5 flippers get heads at once}&.75&3&$\infty$&2&$\infty$&no\\
&&.95&2&$\infty$&2&$\infty$&no\\\hline
\end{tabular}
\vspace{10pt}

The table illustrates a general difference between relatively fine-grained questions (such as \textit{exact outcome} or \textit{outcome shape}) and more coarse-grained ones (such as \textit{how many total tails}, \textit{how long until over}, and \textit{how many end together}). When the topic of our knowledge aligns with a coarse-grained question, we will generally know more relative to that question than we know relative to a more fine-grained question: for example, we know more about how many tails there will be relative to \textit{how many total tails} than we do relative to \textit{exact outcome} or \textit{outcome shape}. But this increase in knowledge comes at a cost, since relative to a coarse-grained question we will know very little about the many topics that are orthogonal to that question: for example, we know little about how many tails there will be relative to \textit{how long until over} or \textit{how many end together}. These two facts share a common explanation. By treating all worlds that agree on the answer to a coarse-grained question as equally normal, we make it easier to exceed $t$ as we add probabilities along the normality order while staying within a relatively restricted class of answers to that question, thus generating a lot of knowledge about that question. But in doing this, we will be including some worlds amongst the relatively normal ones in which things unfold in the least probable way they might with regard to some orthogonal subject matter.

\section{Inductive knowledge from instrument readings}\label{sec:clock}

We often learn about the values of continuous quantities like weight and temperature by measuring them using less than perfectly reliable scales, thermometers, and so on. Modeling such knowledge requires two generalizations of the present framework. One concerns cases where there are a continuum of possible bodies of evidence, so $P(E)=0$ for some $E\in \mathcal{E}$. To handle such cases, we relax the requirement that evidential probabilities are always the result of conditioning a prior probability distribution on your evidence. Instead, we directly associate every $E\in \mathcal{E}$ with a probability distribution $P_E$ over $E$. 
A second problem concerns cases where $Q$ has a continuum of answers all of which have evidential probability 0, yet we want to allow for non-trivial inductive knowledge. Here the natural solution is to generalize the operative notion of probability to \emph{probability density}; see appendix~\ref{app:density} for the technical details. Our example in this section will illustrate both of these issues. Again, we will show that the framework allows us to derive (from purely probabilistic considerations) models of agents' knowledge and beliefs that have been defended in the literature on independent grounds. 

Consider the kind of case made famous by Williamson \cite{Will14}. You are going to glance at an unmarked modernist clock, with only an hour hand. $S=[0,2\pi)\times [0,2\pi)$, where $\langle x,y\rangle$ is a state in which the hand's orientation is $x$ (so that, e.g., $\frac{\pi}{2}$ represents 3-o-clock) and its apparent orientation when you look at it is $y$. We assume that, before looking at the clock, you have no idea how it will look or what time it is; after looking at the clock, your evidence is exhausted by how it appeared. That is, $\mathcal{E} = \{S\}\cup\{\{s\in S: \pi_2(s)=y\}: y\in [0,2\pi)\}$, and $P_S$ is uniform concerning both the hand's real and apparent orientations, in the sense that, for any interval $I=[a,b]\subseteq[0,2\pi)$, $P_S(\{s: \pi_1(s)\in I\})=P_S(\{s: \pi_2(s)\in I\})=\frac{b-a}{2\pi}$. Since your evidence after seeing the clock (i.e., that the apparent orientation was $y$) had prior probability 0, your new evidential probabilities cannot be given by conditioning your prior evidential probabilities on your new discovery.  Moreover, if the question is how the hand is oriented -- i.e., if $Q=\{\{s\in S: \pi_1(s)=x\}: x\in [0,2\pi)\}$ -- then, since your eyesight is imperfect, each of its answers will still have probability 0 after looking at the clock. The case thereby illustrates both issues described in the last paragraph. 

To allow for non-trivial inductive knowledge concerning the position of the hand, we must modify \textsc{normality as likeliness} so that evidentially accessible worlds can differ in normality after looking at the clock. Fortunately, there is a natural way to do this. The key observation is that, after looking, not all \emph{intervals} of orientations are on a par -- their probabilities are no longer given merely by their length. Their probabilities are given instead by a non-uniform \emph{probability density} function, a ``bell curve'' centered on the apparent orientation $y$. The area under this curve between two points gives the probability that the hand's true orientation is in that interval. This fact suggests modifying the definition of $\succcurlyeq$ as follows: rather than ordering worlds according to the respective \emph{probabilities} of their answers to $Q$, we can instead order them by the respective \emph{probability densities} (i.e., heights of the ``bell curve'') of their answers to $Q$. The formal details are given in appendix~\ref{app:density}. 

The normality structures generated in this way determine epistemic and doxastic accessibility relations of the same kind that have been defended in the literature. The agent's beliefs about the hand's orientation will be characterized by a non-trivial interval centered on its apparent orientation. Their knowledge will also be characterized by such an interval, in a way that may or may not always leave a ``margin for error'', depending on whether we adopt a Williamsonian or a Stalnakerian definition of epistemic accessibility.\footnote{Williamson \cite{Will14} and Stalnaker \cite{Stal15} respectively defend such models of our knowledge about the unmarked clock. In \cite{GoodSaloEpis}, we discuss how other models in the literature arise from various combinations of definitions of epistemic accessibility and claims about the comparative normality of the relevant possibilities.} We believe that the present framework is strongly recommended by its ability to vindicate natural and anti-skeptical models of our knowledge in cases of this kind.

\section{Conclusion}

In this paper we have offered a new framework for modeling inductive knowledge and (rational) belief using resources congenial to philosophers in the Bayesian tradition. We did so by showing how the relations of comparative normality that have recently been used to model knowledge and belief can themselves be analyzed in probabilistic terms. The framework offers a unified account of our knowledge about chancy processes (section \ref{sec:flipping}), lawful regularities (section \ref{sec:laws}), and imprecisely measured quantities (section \ref{sec:clock}). By positing a certain kind of context-sensitivity in ``know'' and ``believe'', it offers a way of avoiding inductive skepticism while maintaining that only highly probable propositions are ever known or rationally believed (section \ref{sec:racing}).  An urgent question for further research is how the contextually-supplied question that features in the probabilistic analysis of normality is determined.\footnote{Note that relative to the question \emph{is it true that $p$}, knowing that $p$ requires only that $p$ is true and has a high enough evidential probability. This might be considered objectionable for familiar reasons to do with Gettier cases \cite{Gett63}. If so, that would be one reason to deny that such questions are supplied by any context -- although see \cite{HolgMSconstraint} for arguments that there are contexts in which ``knowledge'' is this easy to come by.}



\section*{Acknowledgements}

Thanks to Cian Dorr and three anonymous referees for TARK for comments on a draft of this material, and to Kevin Dorst, John Hawthorne, Ben Holgu\'in, and Harvey Lederman for very helpful discussion. Special thanks to Dorst for help with the coding required to calculate the values in the table in section \ref{sec:racing}. 
\appendix

\section{Primitive evidential accessibility}\label{app:Generalizations}

Rather than modeling worlds as state/set-of-state pairs, we could treat them as unstructured points, and explicitly specify an evidential accessibility relation on them. This no longer allows us to model the notion of discovery, but it avoids the presupposition that there is a principled way of factoring worlds into state/set-of-state pairs. Moreover, it allows us to model evidential accessibility as an arbitrary reflexive relation. To handle cases where evidential accessibility is not an equivalence relation, we will need to relativize relations of comparative normality to a reference world -- the world whose evidential probabilities are being used to assess the comparative normality of two other worlds.\footnote{\cite{Will00} argues that evidential accessibility is not an equivalence relation; \cite{Lewi96} and \cite{Stal19} maintain that it is. \cite{Cart19} and \cite{LoetMS} argue for a kind of world-relativity of (non-comparative) normality; see \cite{GoodSaloEpis} for discussion in the context of comparative normality; ordering semantics for counterfactuals is a formal precedent, \emph{cf.} \cite{Lewi73}.}

Let a \emph{relativized normality structure} be a tuple $\langle W,R_e,\succcurlyeq,\ggcurly\rangle$ such that:
\begin{enumerate}
    \item $W$ is a non-empty set,\label{annoying1}
    \item $R_e: W\to\mathcal{P}(W)$ such that $w\in R_e(w)$ for all $w\in W$,\label{annoying2}
    \item For each $w\in W$, $\succcurlyeq_w$ is a preorder on $W$,
    \item For each $w\in W$, $\ggcurly_w$ is a well-founded relation on $W$ such that, for any worlds $w_1,w_2,w_3,w_4$:\begin{enumerate}
        \item If $w_1\ggcurly_w w_2$, then $w_1\succcurlyeq_w w_2$;
        \item If $w_1\succcurlyeq_w w_2 \ggcurly_w w_3 \succcurlyeq_w w_4$, then  $w_1 \ggcurly_w w_4$.
        \end{enumerate}
\end{enumerate}
We define $R_b(w)$ and $R_k(w)$ as in section~1, replacing $\succcurlyeq$/$\ggcurly$ with $\succcurlyeq_w$/$\ggcurly_w$.

 Now let a \emph{worldly probability structure} be a tuple $\langle W,R_e,Q,P,t\rangle$ such that:
\begin{enumerate}
    \item $W,R_e$ satisfy \ref{annoying1} and \ref{annoying2} in the definition of a relativized normality structure,
    \item $Q$ is a partition of $W$,
    \item $P$ is a probability distribution over $W$ such that $P(q|R_e(w))$ is defined for all $q\in Q$ and $w\in W$,
    \item $t\in [0,1]$.
\end{enumerate} 
Let $\lambda_w(v):=P([v]_Q|R_e(w))$ and $\tau_w(v):=P(\{u: v\succcurlyeq_w u$ and $u\in R_e(w)\}|R_e(w))$. With these world-relative notions of likeliness and typicality in hand, we can now define $v\succcurlyeq_w u:=\lambda_w(v)\geq \lambda_w(u)$ and  $v\ggcurly_w u:=1-\frac{\tau_w(u)}{\tau_w(v)} \geq t$. It is easy to verify that these definitions yield a relativized normality structure that also obeys \textsc{threshold} (reformulated with $R$ in place of the now ill-defined $\mathcal{R})$.

\section{Probability densities}\label{app:density}

Let a \emph{density structure} be a tuple $\langle S,\mathcal{E},W,P,Q,m,f,t\rangle$ such that:\footnote{It would also be natural to require that, where possible, $P$ behaves as if it were the result of conditioning a prior on the agent's evidence -- i.e. that, for all $E,E'\in \mathcal{E}$, if $P_E$ is defined on $E\cap E'$, then $P_E(\cdot|E\cap E')=P_{E'}(\cdot|E\cap E')$.}
\begin{enumerate}
    \item $S$, $\mathcal{E}$, $W$, $Q$, and $t$ are as in a probability structure and $t>0$,
    \item For all $E\in \mathcal{E}$:
    \begin{enumerate}
        \item $P_E$ is a probability distribution over $E$,
        \item $P_E(q)=0$ for all $q\in Q$,
    \end{enumerate}  
    \item $m: Q\to \mathbb{R}$ (the \emph{measuring function}),
    \item $f:\mathcal{E}\to \mathbb{R}^{\mathbb{R}}$ such that $f_E$ is the \emph{density of $P_E$ relative to $m$}.
\end{enumerate}
The intuitive idea behind a probability density function is that of a curve the area under which gives the associated probabilities. So, in particular,   $\int_a^bf_E(x)dx=P_E(\bigcup\{q\in Q: m(q)\in[a,b]\})$. A formal characterization of $f$ is given in a footnote; the role of $m$ will be illustrated in appendix~\ref{app:dese}.\footnote{Formally, $f_E$ is \emph{the density of $P_E$ with respect to the reference measure $\mu'$} (defined as usual in terms of the Radon-Nikodym derivative), where  $\mu'(p)=\mu\{m(q):q\subseteq p\}$ for all $p$ on which $P_E$ is defined, and $\mu$ is the Lebesgue measure on $\mathbb{R}$.}

To understand this definition, let us return to the unmarked clock. As described in section~\ref{sec:clock}, $P$ satisfies clause 2 (where $Q$ is \emph{what is the hand's orientation}). Let $m([s]_Q)=\pi_1(s)$ -- i.e., it maps answers to $Q$ (understood as sets of states) to corresponding real numbers in $[0,2\pi)$. $f_S(x)=\frac{1}{2\pi}$, the constant function. By contrast, $f_{\{s\in S:\pi_2(s)=y\}}$ -- the probability density function determining your evidential probabilities after discovering that the hand's apparent orientation is $y$ -- will be a `bell curve' centered on and symmetric around $y$ (e.g. a `wrapped normal distribution'). In cases like this, rather than generating a normality order via \textsc{normality as likeliness} from a probability function, we instead generate it from this probability \emph{density} function. Let $d(w)$ (the \emph{density} of $w$) be $f_{\pi_2(w)}(m([\pi_1(w)]_Q))$, and define being at least as normal as follows:
\begin{itemize}
\item[]\textsc{normality as density}: $w\succcurlyeq v:=d(w)\geq d(v)$ and $v\in R_e(w)$
\end{itemize}
As before, $\ggcurly$ is defined by \textsc{sufficiency}.\footnote{Requiring that $t>0$ ensures that $\ggcurly$ is well-founded; for an illustration of why this is needed, see $d^I$ in appendix~\ref{app:dese}.}

To see this definition in action, we will consider a modification of the clock case that allows us to work with familiar Gaussian probability density functions (also known as `normal distributions'). We do so by considering a continuous quantity whose values are not confined to $[0,2\pi)$, such as the difference in weight of two objects. Suppose $Q$ is the question \emph{how much heavier is this apple than this orange}, and our scale reads $\mu$ grams. Suppose, as a first approximation, that our evidential probabilities are now characterized by a Gaussian probability distribution over weight-difference in grams, with mean $\mu$ and standard deviation $\sigma$. $m$ will be the function from answers to $Q$ to corresponding real numbers of grams and $f_E(x)=\frac{1}{\sigma\sqrt{2\pi}}e^{-\frac{1}{2}(\frac{x-\mu}{\sigma})^2}$, where our evidence $E=\{\langle x,\mu\rangle:x\in\mathbb{R}\}$. (We model states as ordered pairs of actual and measured values of the quantity, as before.) By setting $t=.9545\dots$, we predict that $\mathcal{R}_b(\langle\langle x,\mu\rangle,E\rangle)=\{\langle x',\mu\rangle:|x'-\mu|\leq 2\sigma\}$ for all $x\in \mathbb{R}$: we will believe that the scale reads $\mu$ grams and that the true weight-difference is within two standard deviations of that. Predictions concerning knowledge depend on which of the two clauses for $R_k$ are adopted, but in either case our knowledge will be non-trivial (and will coincide with what we believe when the scale is perfectly accurate -- i.e., when $x=\mu$). This model (simplified from \cite{Good13}) shows how, at least in favorable circumstances, we can see familiar uses of confidence intervals in inferential statistic as corresponding to non-trivial knowledge and (rational) beliefs about the values of imprecisely measured quantities.\footnote{In the case of belief, our models correspond to the ``minimum likelihood'' method for generating confidence intervals (which yields the shortest possible intervals that have probability $t$). The idea of ordering possibilities by probability density for this purpose goes back to \cite{NeymPear31}. This procedure has been criticized in the case of asymmetric distributions; in particular, in the case of $\chi^2$ distributions, where it yields different verdicts from ordinary $\chi^2$ tests \cite{GibbPrat75, RadlAlf75, Kuli08}. We lack the space to address these criticisms here, except to note that the \textbf{Decay} example in appendix~\ref{app:dese} illustrates why we find appealing the distinctive predictions that the minimum likelihood approach makes in the case of certain asymmetric distributions.}  

Our view is not that normality relations are \emph{always} determined by density structures. Sometimes they are determined by probability structures. It depends on whether the answers to $Q$ have positive probabilities or only probability densities. When answers differ in this regard, we advocate a hybrid approach, with $\succcurlyeq$ determined by $\lambda$ among pairs of answers one of which has positive probability and by $d$ among pairs of answers with well-defined probability densities. Another generalization of density structures is also needed to handle multidimensional probability densities. In the case of $n$ dimensions, we will then have $f_E:  \mathbb{R}^n\to \mathbb{R}$. This generalization is needed when the relevant quantity is multidimensional (e.g., where a dart will land on a dartboard) and/or when $Q$ concerns the outcomes of multiple independent noisy measurements of a given quantity (which will be modeled as a vector of real numbers). 

\section{Normality \emph{de dicto} and \emph{de se}}\label{app:dese}

In the main text, we modelled a question as a partition of $S$. However, some natural questions cannot be modelled in this way, because they don't merely concern the state of the world. In \textbf{Flipping for Heads}, we might, for example, wonder \textit{how many \underline{more} times a coin will be flipped} -- two worlds in which it is flipped a total of 5 times can differ on the answer to this question, because in one it is part of your evidence that the coin has already been flipped (only) 2 times while in another it is part of your evidence that it has been flipped 3 times. Following \cite{Lewi79}, we will think of these as \textit{de se} questions, concerning not only the history of the world but also your place in it. Formally, we implement this idea by modeling $Q$ as a partition of $W$, as in appendix~\ref{app:Generalizations} -- recall that members of $W$ should be thought of as \emph{centered} worlds, two of which can agree on the complete history of the world while disagreeing on your evidence (because they concern your evidential situation at different times in that history).

Unlike appendix~\ref{app:Generalizations}, for present purposes we may keep $P$ defined on subsets of $S$ -- this is natural in cases where we want $P$ to conform to the prior objective chances which are (plausibly) only defined over histories of the world (i.e., states). Since $P$ is now not defined on answers to $Q$ (these being sets of worlds), we need to modify the definition of $\lambda$. For $q\in Q$, we let $q_E=\{\pi_1(w):w \in q \text{ and } \pi_2(w)=E\}$; then $Q_E:=\{q_E:q\in Q\}$ is a partition of $E$. So we can redefine $\lambda(\langle s,E\rangle)$ as $P([s]_{Q_E}|E)$.\footnote{This definition is a proper generalization of the previous one since, if we start with a partition $Q^S$ of $S$ and use it to generate a partition $Q^W$ of $W$ in the obvious way, defining likeliness as above using $Q^W$ yields the same results as the original definition of likeliness using $Q^S$.}

We will illustrate this generalization by showing how it allows us to model the following case (also discussed in \cite{GoodSaloEpis}):
\begin{quote} \textbf{Decay}: A radioactive atom is created; eventually, it will decay. The average time for an atom of this isotope to decay is one year.
\end{quote}
Assume that the agent's evidence entails that this is the set-up, that they are keeping track of time, and that, when the atom decays, they will be alerted of this fact. So states can be modeled as positive real numbers, specifying how many years from its creation it will be before the atom decays, and possible bodies of evidence (in which the atom has yet to decay) are intervals $(t,\infty)$ for positive real $t$. 

What should the agent believe about how many years after its creation the atom will decay? A natural thought is that, at every time $t$ before the atom decays, the agent's doxastic possibilities will be characterized by the interval  $[t+a,t+b]$, for some $0<a<1<b$. (Compare other events that tends to happen unexpectedly, like when you will next have an urge to sneeze, or when you will next get an email from an intermittent correspondent: you think they won't happen in the next second, you think they will have happened within a decade, and the doxastically possible times form an interval, at least to a first approximation.) 
To ensure that the shape of this interval is the same at every time, we model the case using the \textit{de se} question \textit{how long after the current time will the atom decay}. To predict that the agent believes that the atom will take at least $a$ years to decay, we need to choose the measuring function $m$ of our density structure to reflect the fact that extremely short times to decay can be as far from average, in the relevant sense, as extremely long times to decay. 

We will now make these ideas precise using the framework of density structures from appendix~\ref{app:density} (generalized to allow \emph{de se} questions). Let $S=\mathbb{R}^+$,  $\mathcal{E}=\{(t,\infty): t\geq 0\}$, and let $P_{(t,\infty)}$ be given by the objective chances at $t$ (which are entailed by the agent's evidence at $t$, since it entails the setup and the time). $Q=\{q_r:r\in \mathbb{R}^+\}$, where $q_r=\{\langle t',(t,\infty )\rangle: t'-t=r\}$. But this does not yet specify a density structure, since it is compatible with different choices of measuring functions $m$ and corresponding probability density assignments $f$. 

How should we associate answers to $Q$ with real numbers? The simplest choice is the \emph{index} function $m^I$, where $m^I(q_r)=r$. Given this choice, the corresponding density $f^I_E(x)=e^{-x}$ (for all $E\in \mathcal{E}$, so the agent's beliefs about how much longer it will be before the atom to decays don't change if they see the atom hasn't decayed yet), from which it follows that  $d^I(\langle t',(t,\infty)\rangle )=e^{t-t'}$, since $[\langle t',(t,\infty)\rangle]_Q=q_{t'-t}$. Since the shortest lengths of time until decay correspond to the highest densities, it is always doxastically possible that the atom will decay arbitrarily soon. 



\begin{figure}[h]
\centering
\parbox{.05\linewidth}{$f^I_{(t,\infty)}(x)$}%
\hspace{.05\linewidth}%
\parbox{.25\linewidth}{\includegraphics[width=\linewidth]{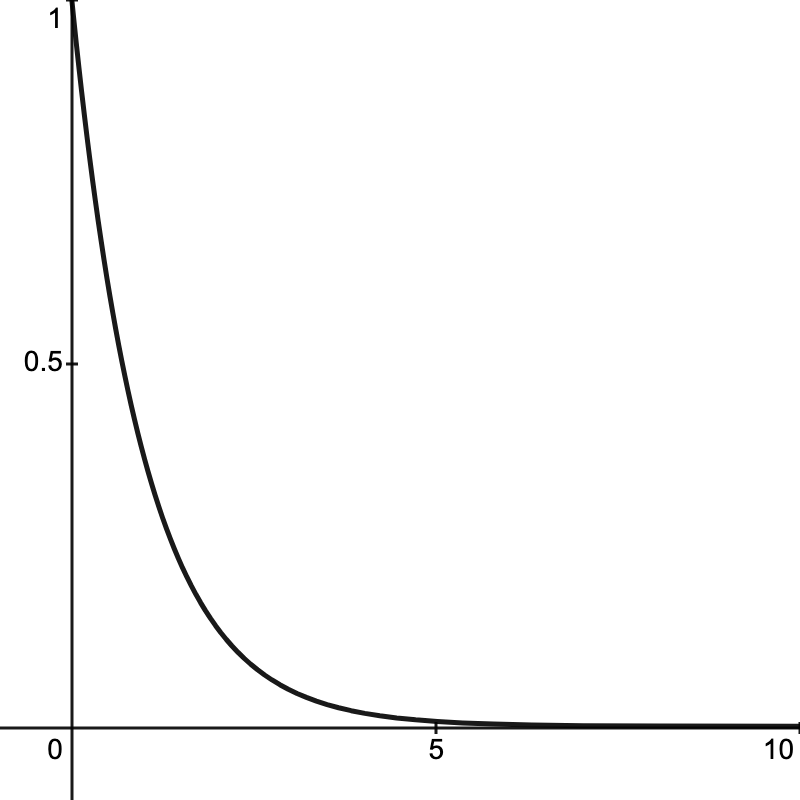}}
\hspace{.1\linewidth}
\parbox{.05\linewidth}{\subcaption*{$d^I(q_r)$}}%
\hspace{.02\linewidth}%
\parbox{.25\linewidth}{\includegraphics[width=\linewidth]{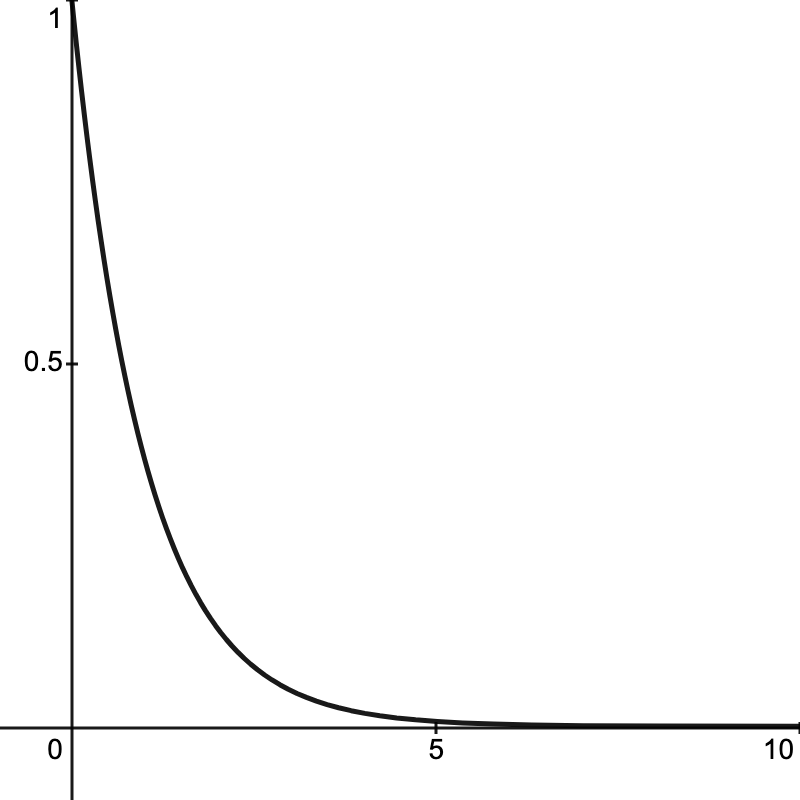}}\\

\hspace{.25\linewidth}\parbox{.15\linewidth}{$x$} \hspace{.25\linewidth}\parbox{.15\linewidth}{$r$}


\end{figure}

To avoid this prediction, we must choose a different measuring function. And there is a natural alternative: the \emph{logarithmic} measuring function $m^{\ln}$, where $m^{\ln}(q_r)=\ln(r)$. This choice corresponds to a different density function $f^{\ln}_E=e^{x}e^{-e^x}$ (for all $E\in \mathcal{E}$), from which it follows that $d^{\ln}(\langle t',(t,\infty)\rangle )$ = $(t'-t)e^{t-t'}$.\footnote{
By definition, $f_E^{\ln}(x)$ is such that 
    \begin{align*}
    \int_{-\infty}^{y} f^{\ln}_E(x)  & = P_E\Big(\bigcup\big\{q_r: m^{\ln}(q_r)\in(-\infty,y)\big\}\Big)\\
                                    & = \int_0^{e^y} e^{-x}dx \text{\hspace{.5cm}      (for all }E\in\mathcal{E})\\
                                    & = -e^{-e^y} + 1
    \end{align*} Differentiating both sides with respect to $y$ then yields $f_E^{\ln}(x)=e^{x}e^{-e^x}$. So $d^{\ln}(\langle t',(t,\infty)\rangle )=f_{(t,\infty)}^{\ln}(m^{\ln}([\langle t',(t,\infty)\rangle]_Q))=f_{(t,\infty)}^{\ln}(\ln(t'-t))= (t'-t)e^{t-t'}$.
    } 
Intuitively, this choice of measuring function amounts to thinking of intervals between the present time and the time of decay in terms of the \textit{order of magnitude} of their duration. In addition to being mathematically natural, this choice is psychologically well-motivated, as there is a large body of psychometric research showing that we perceive temporal duration in this way; see \cite{GCW84}. 

\begin{figure}[h]
\centering
\parbox{.05\linewidth}{$f^{\ln}_{(t,\infty)}(x)$}%
\hspace{.05\linewidth}%
\parbox{.25\linewidth}{\includegraphics[width=\linewidth]{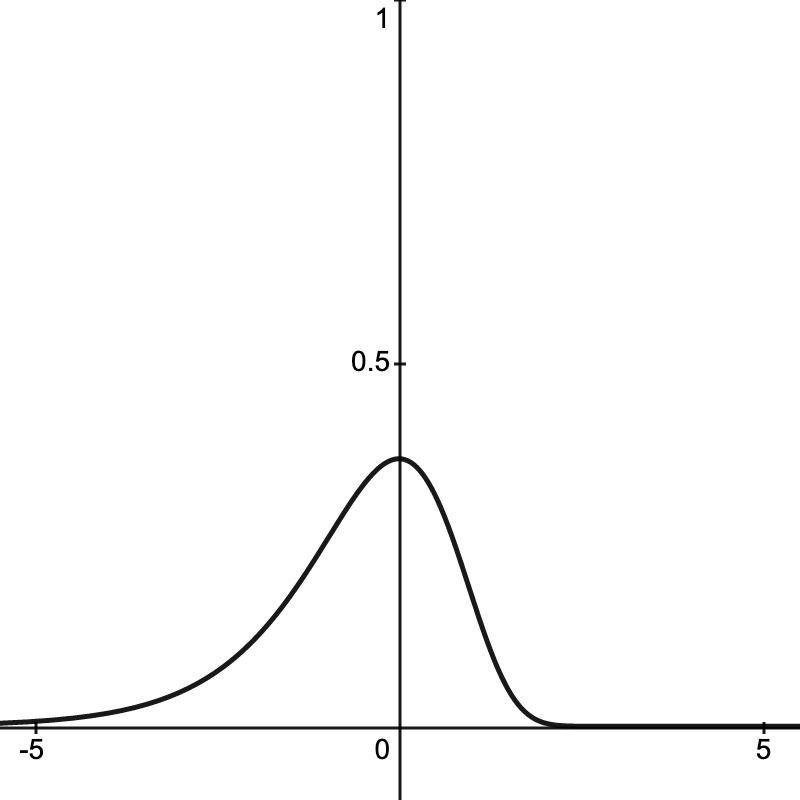}}
\hspace{.1\linewidth}
\parbox{.05\linewidth}{\subcaption*{$d^{\ln}(q_r)$}}%
\hspace{.02\linewidth}%
\parbox{.25\linewidth}{\includegraphics[width=\linewidth]{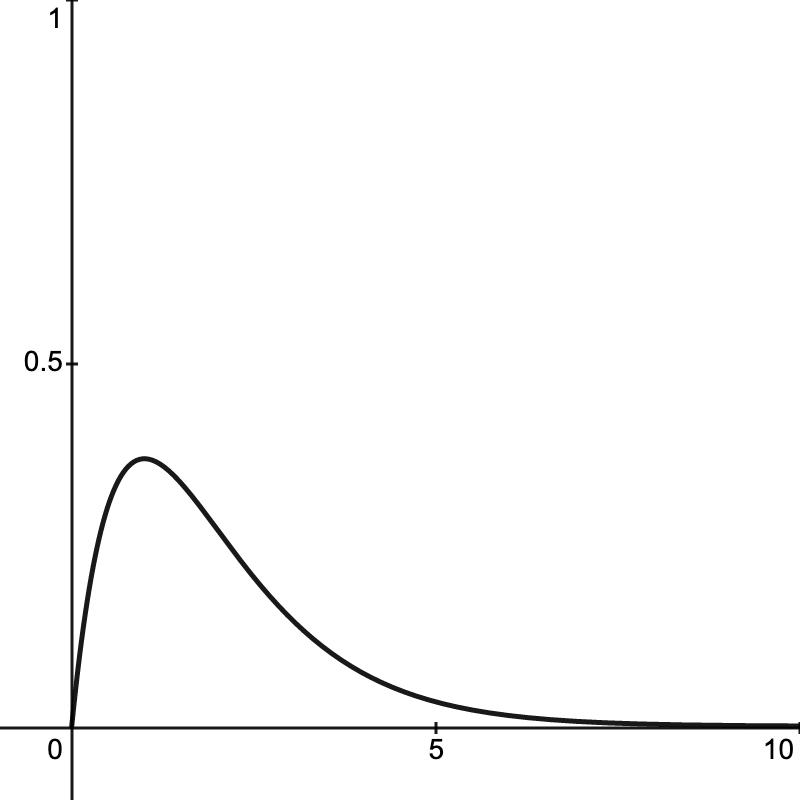}}\\

\hspace{.25\linewidth}\parbox{.15\linewidth}{$x$} \hspace{.25\linewidth}\parbox{.15\linewidth}{$r$}


\end{figure}

Unlike $d^I(q_r)$ (which has its highest values as $r$ approaches $0$), $d^{\ln}(q_r)$ approaches $0$ as $r$ approaches $0$. This means that, for $r< 1$, smaller values of $r$ make for increasing abnormality, and (centered) worlds in which the atom is going to decay extremely soon are hence doxastically inaccessible. More generally, the doxastically possible answers to $Q$ are those whose value of $d$ falls above a certain horizontal line (which depends on the probability threshold $t$); inspection of the graph of $d^{\ln}$ above shows that, as desired, this will always be an interval that excludes both possibilities in which the atom decays immediately and ones in which it won't decay for a very long time.


This model combines two ideas: the appeal to a \textit{de se} question, and the appeal to a logarithmic measuring function. One might wonder whether the first of these is really necessary. Couldn't we have achieved the same effect by using a logarithmic measuring function on the \textit{de dicto} question $Q^*$: \textit{how long after its creation does the atom decay}? The answer is ``no'' -- doing so achieves the same effect only at the moment the atom is created. This is because, unlike the probability densities of answers to $Q$, the probability densities of answers to $Q^*$ change as new evidence becomes available: when the atom hasn't decayed 9 months after its creation, the probability that it will decay within the \textit{next} year remains unchanged, but the probability that it will decay within its \textit{first} year clearly decreases. (In particular, after a year, the answer to $Q^*$ that implies that the atom will decay immediately will have the highest probability density even with a logarithmic measuring function; so possibilities in which the atom decays immediately will then be doxastically accessible.) Intuitively, it is unsurprising that a \textit{de se} question is needed for the desired dynamical behavior in \textbf{Decay}. The idea that one is always entitled to believe that the atom isn't about to decay relies on there being something particularly odd about it decaying \textit{right now} or \textit{very soon}; but that oddity is clearly tied to \textit{de se} notions, and cannot be articulated without them.


\bibliographystyle{eptcs} 
\bibliography{Bibliography}

\end{document}